\documentclass{mem}
\usepackage{natbib}\usepackage{txfonts}\usepackage{balance}
\usepackage{graphicx}
\usepackage[a4paper,breaklinks,dvipdfm]{hyperref}
\idline{75}{282}
\begin{document}
\def\teff{$T\rm_{eff }$}
\def\kms{$\mathrm {km s}^{-1}$}

\title{
Formation Mechanisms of IMBH in Globular Clusters
}

   \subtitle{}

\author{M. \ Giersz\inst{1}, \ N. \ Leigh\inst{2}, \ A. \ Hypki\inst{3}, \ A. \ Askar\inst{1}
\and \ N. \ L\"{u}tzgendorf\inst{4}}

\institute{
Nicolaus Copernicus Astronomical Centre, Polish Academy of Sciences, ul. Bartycka 18, 00-716 Warsaw, Poland
\email{mig@camk.edu.pl}
\and
Department of Astrophysics, American Museum of Natural History, Central Park West and 79th Street, New York, NY 10024
\and
Leiden Observatory, Leiden University, PO Box 9513, NL-2300 RA Leiden, the Netherlands
\and
Space Telescope Science Institute, 3700 San Martin Drive, Baltimore, MD 21218, USA
}

\authorrunning{Giersz et al. }

\titlerunning{IMBH in Globular Clusters}

\abstract{ 
We very briefly discuss proposed in the literature possible scenarios
for intermediate mass black holes formation in globular clusters. We also discuss the results of the MOCCA simulations of about 2000 models (BigSurvey)
regarding the distribution of events connected with electromagnetic and
gravitational radiations, namely: mass transfer on IMBH, collisions and mergers
with IMBH and mergers with IMBH due to gravitational radiation. The rates of these events are very small, so their observation is very improbable.}

\maketitle{}

\section{Introduction}

Intermediate mass black holes (IMBH) are thought to be a missing link between
stellar-mass black holes (BH) and supermassive BHs (SMBH). There are many
theoretical arguments in favor of the formation of IMBHs in the centers of
globular clusters (GCs) \citep[e.g.][and reference therein]{Lutzgendorfetal2013}.
Observational confirmation of the existence of IMBHs would have an important
impact on a number of open astrophysical problems related e.g. formation of
SMBHs and their host galaxies, origins of ultraluminous X-ray sources in nearby
galaxies, and detection of gravitational waves (GR). GCs are thought to be a
natural site for IMBH formation. The proximity of GCs makes possible to
directly observe kinematic and structural imprints of IMBHs. There are a lot
of observations indirectly suggesting the presence of IMBHs in GCs in nearby
galaxies or Milky Way. They are based on the detection of strong X-ray or radio
emissions at of-centre positions in distant galaxies, not confirmed X-ray or/and
radio emissions in some Galactic GCs or on kinematic and spatial structure of
central parts of GCs. Up to now, there is no firm observational confirmation of
IMBH presence in GCs.

\section{IMBH formation mechanisms}

In the literature, so far, there were proposed five possible scenarios for IMBH
formation in GCs:
\begin{enumerate}
\item Direct collapse of very massive Population III stars \citep{MadauRees2001}.
The very first generation of stars must have formed out of unmagnetized,
metal-free gas. Stars were formed at so called mini halos of mass about $10^5
M_{\odot}$ at redshift about 20. The Population III stars are formed with
initial mass function (IMF) extremely top heavy, so stars with masses larger
than $200 - 300 M_{\odot}$ can be formed. After a few Myr of evolution the most
massive Population III stars ended up as BHs, IMBH seeds, of masses larger
than $150 - 200 M_{\odot}$, loosing only small fraction of mass \citep{FryerWH2001}. However, one can ask a question relevant for the presence
of IMBHs in GCs. How could such extremely massive BHs become members of GCs
consisting of Population II stars?
\item Runaway merging of very massive main sequence stars (MS) stars in dense
young star clusters was first discussed by \citet{Portegiesetal2004}. In this
scenario there are needed very tailored initial conditions to initiate a runaway
merging process. Time scale of the mass segregation of the most massive stars
(about $100 M_{\odot}$) has to be shorter than the evolution time-scale for
those stars, otherwise massive stars will evolve before they will start to
collide. The velocity dispersion in the collapsed star cluster cannot be larger
than a few hundred $km/s$, otherwise collisions would disrupt stars.
Frequent collisions between stars lead to the formation of very massive stellar type
objects like MS. Such star will quickly evolve and become a massive BH and an IMBH seed.
\item Accretion of the residual gas on stellar mass BHs formed from the first
generation stars was proposed by \citet{Leighetal2013}. Actually, the main
purpose of this model was to explain multiple stellar populations in GCs.
Interactions between interstellar medium remaining after formation of the first
generation stars and BHs formed from the most massive first generation stars is
associated with substantial mass accretion onto stellar mass BHs. This leads to
substantial increase of the BH masses and speedup of their mass segregation and
finally, to formation of a very massive BHs, IMBH seeds. But we should be aware
that the residual gas removal takes a few Myr and it is comparable with BH
formation time-scale, so probably there is not enough time and not enough
residual gas to build very massive BHs, unless the residual gas is very dense
and can stay in the cluster longer.
\item A more observational than theoretical scenario, which is based on the
observed relation between the BH mass and the velocity dispersion of its host
galaxy \citep[e.g.][and references therein]{Gultekin2009}. Extrapolating this
relation to the velocity dispersions characteristic of Galactic GCs, one expects
to find central BHs with masses similar to those characteristic of IMBHs.
\item And finally, the new, just recently presented, scenario based on results
of MOCCA simulations of evolution of dense stellar systems
\citep{Gierszetal2015}. In this scenario IMBH is formed because of buildup of BH
mass solely due to mergers in dynamical interactions and mass transfers in
binaries. In this scenario, in contrast to the scenarios presented above, there
is no need for any special conditions to initiate the process of IMBH mass
buildup. However, the process of IMBH formation is highly stochastic. The larger
the initial cluster concentration, the earlier, faster and with higher probability
an IMBH will form. It is worth to stress that IMBH formation does not strongly
depend on details of mass accretion onto BH and the detailed structure of a star
after physical collision with another star.
\end{enumerate}

\begin{table*}[t!]
\begin{minipage}{150mm}
  \caption{Initial conditions for the BigSurvey of the MOCCA simulations.}
  \label{tab:models}
\begin{tabular}{|l|c|c|c|c|c|c|c|c|}
\hline
N & $R_t$ (pc) & $R_t/R_h$ & $W_0$ & Z & $f_b$ & $a_{max}$ (AU) & IMF & Kicks (km/s) \\
\hline
$4.0x10^4$ & 30, 60 & 25, 50  & 3.0, 6.0 & 0.001, 0.005 & 0.05, 0.1 &     100    & IMF2 & 265 \\
             &   120  & filling &    9.0   &     0.02     & 0.3, 0.95 &   Period &     & Fallback \\
\hline
$1.0x10^5$ & 30, 60 & 25, 50  & 3.0, 6.0 & 0.001, 0.005 & 0.05, 0.1 &     100    & IMF2 & 265 \\
             &   120  & filling &    9.0   &     0.02     & 0,3, 0.95 &   Period &       &Fallback \\
\hline
$4.0x10^5$ & 30, 60 & 25, 50  & 3.0, 6.0 &    0.001     & 0.05, 0.1 &     100    & IMF2 & 265 \\
             &   120  & filling &    9.0   &              &   0.95    &   Period &       &Fallback \\
\hline
$7.0x10^5$ & 30, 60 & 25, 50  & 3.0, 6.0 & 0.0002, 0.001, 0.005 & 0.05, 0.1 &     100   & IMF2 & 265 \\
             &  120   & filling &    9.0   &  0.006, 0.02 & 0,3, 0.95 &   Period &       &Fallback \\
\hline
$1.2x10^6$ & 30, 60 & 25, 50  & 3.0, 6.0 &     0.001    & 0.05, 0.1 &     100    & IMF2 & 265 \\
             &  120   & filling &    9.0   &              &    0.95   &   Period &       &Fallback \\
\hline
\end{tabular}
\vskip 0.2cm
 \textit{Notes:}
BH and NS kicks are the same \citep{Hobbsetal2005}, except the case of mass fallback \citep{Belczynskietal2002}. IMF2 - two segmented initial mass function \citep{Kroupa2001}, Period - binary period distribution \citep{Kroupa1995}. $f_b$ - binary fraction, $N$ - number of stars, $R_t$ - tidal radius, $R_h$ - half-mass radius, $W_0$ - King model parameter, $Z$ - cluster metalicity, $f_b$ - binary fraction, $a_{max}$ - maximum value for semi-major axis (distribution uniform in $log(a)$, filling - tidally filling model).
\end{minipage}
\end{table*}

\section{Method and models}

The MOCCA (MOnte Carlo Cluster simulAtor) code used for the star cluster simulations presented here is a numerical simulation code based on H\'{e}non's implementation of the Monte Carlo method \citep{Henon1971} to follow the long term evolution of stellar clusters. This method was substantially developed and improved by Stod\'{o}{\l}kiewicz in the early eighties \citep{Stodolkiewicz1986} and later by Giersz and his collaborators \citep[and reference therein]{Gierszetal2013}. The method can be regarded as a statistical way of solving the Fokker-Planck equation. The MOCCA code has been extensively tested against the results of N-body simulations of star clusters containing from thirty thousands up to one million stars \citep[][and references therein]{Gierszetal2013,Heggie2014,Gierszetal2015,Wangetal2016,Mapelli2016}. The agreement between these two different types of simulations is excellent.  This includes the global cluster evolution, mass segregation time scales, the properties of primordial binaries (energy, mass and spatial distributions), and the numbers of retained neutron stars (NS) and BHs.

For the new scenario of IMBH formation we analyse results from 1950 star cluster
models that were simulated using the MOCCA code (so called BigSurvey). The
simulations are characterized by very diverse parameters describing not only the
initial global cluster properties, but also star and binary properties. The
parameters of these models are listed in Table \ref{tab:models}.

We would like to strongly stress that models for the BigSurvey were not selected
to match the observed Milky Way GCs. Nevertheless, as it is shown in
\citet{Askaretal2016} the agreement with the observational properties of
observed Galaxy GCs \citep[][updated 2010]{Harris1996} and modeled ones is quite
good for the cluster absolute magnitude and the average surface brightness.
Despite this agreement, its is important to state that any combination of global
observational properties of GCs cannot be used to distinguish between different
cluster models because there is a strong degeneracy with respect to the initial
conditions. Taking into account the relatively good agreement between the
observed Milky Way GCs and the modeled ones we can assume that the BigSurvey
cluster models are representative for the GC population.

\section{Observational properties of IMBH binaries}

As it is well known, BHs cannot be directly observed because no light can leave them. Nevertheless, we can observe them indirectly by characteristic electromagnetic or gravitational emissions or because of their influence on spatial and kinematic characteristics of star clusters. Here, we will concentrate only on cumulative distributions of events connected with electromagnetic emissions during mass transfer on IMBH from a companion star, electromagnetic emissions during tidal disruption events (TDE) or direct collisions with IMBHs. We also provide cumulative distributions for merger events between IMBHs and compact stars (white dwarfs (WD, NS or BHs) that are induced by gravitational radiation (GR). 

\begin{figure}[]
\resizebox{\hsize}{!}{\includegraphics[angle=270,clip=true]{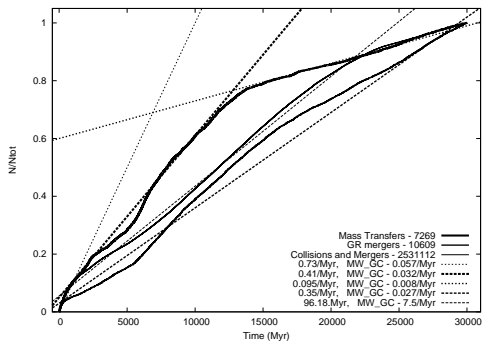}}
\caption{
\footnotesize
The cumulative distributions of GR mergers, mass transfer and merger/collision
events with IMBHs.
Line descriptions and total number of events are given in the figure inserts.
}
\label{fig:fig1}
\end{figure}
   
On the Fig. \ref{fig:fig1} you can see the cumulative distributions of all GR merger, mass transfer and merger/collision events with IMBH. The total numbers of GR events, mass transfer events and mergers/collisions events are very large and are 10609, 7269 and 2531112, respectively. Those numbers are for models in which IMBH is formed (about $25\%$ of all models in the BigSurvey).  
Roughly, the cumulative distributions can be approximated by straight lines. Assuming that simulated models are representative for population of real GCs we can estimate the rate of occurrence of a particular type of events. These numbers are very small. The largest rate is for mergers/collisions with IMBH. It is about 100 per Myr. Others are less than 1 per Myr. If we scale those rates to the number of Milky Way GCs we will get even smaller values. For mergers/collisions it is of the order of a few per Myr. Moreover, such events have a very short duration, so there is a low probability of observing them. So, we should not be surprised that there is a rather small chance to observe events connected with electromagnetic or gravitational events and why they were not observed yet.  

\section{Conclusions}

The MOCCA code shows clearly its ability to model in a very efficient way a large number (thousands) of GC models. The observation of any events connected with accretion of matter on IMBHs or connected with mergers due to gravitational radiation is very improbable. We will have to have a lot of luck to see one of them. All the simulations data discussed here is a part of the BigSurvey database. This database can be freely accessed. If you are interested in using the data from the BigSurvey in your own research please send an email to Mirek Giersz.

\begin{acknowledgements}
MG, AH and AA were partially supported by Polish National Science Center through grant DEC-2012/07/B/ST9/04412.
\end{acknowledgements}

\bibliographystyle{aa}

\end{document}